\documentclass[aps,prb,showpacs,twocolumn,reprint,superscriptaddress,longbibliography,showkeys]{revtex4-2}
\usepackage{amsmath}
\usepackage{amssymb}
\usepackage{graphicx}
\usepackage{color}
\usepackage{multirow}
\usepackage{dcolumn}
\usepackage{textcomp}
\usepackage{longtable}
\usepackage{makecell}
\usepackage{booktabs}
\usepackage{array} 
\usepackage{ulem}
\usepackage{appendix}
\usepackage{placeins}

\begin{document}


\title{Doping induced multiferroicity and quantum anomalous Hall effect in $\alpha$-In$_2$Se$_3$ thin films}

\author {Zhiqiang Tian}
\affiliation{Key Laboratory for Matter Microstructure and Function of Hunan Province,
Key Laboratory of Low-Dimensional Quantum Structures and Quantum Control of Ministry of Education, School of Physics and Electronics, Hunan Normal University, Changsha 410081, China}

\author{Jin-Yang Li}
\affiliation{School of Physics, Northwest University,  Shaanxi Key Laboratory for Theoretical Physics Frontiers,Xi’an 710127, People’s Republic of China}

\author{Tao Ouyang}
\affiliation{Hunan Key Laboratory for Micro-Nano Energy Materials and Device and School of Physics and Optoelectronics, Xiangtan University, Xiangtan 411105, Hunan, China}

\author{Chao-Fei Liu}
\affiliation{School of Science, Jiangxi University of Science and Technology, Ganzhou 341000, China}

\author{Ziran Liu}
\email{zrliu@hunnu.edu.cn}
\affiliation{Key Laboratory for Matter Microstructure and Function of Hunan Province,
Key Laboratory of Low-Dimensional Quantum Structures and Quantum Control of Ministry of Education, School of Physics and Electronics, Hunan Normal University, Changsha 410081, China}

\author{Si Li}
\email{sili@nwu.edu.cn}
\affiliation{School of Physics, Northwest University,  Shaanxi Key Laboratory for Theoretical Physics Frontiers,Xi’an 710127, People’s Republic of China}

\author{Anlian Pan}
\affiliation{Key Laboratory for Micro-Nano Physics and Technology of Hunan Province, College of Materials Science and Engineering, Hunan University, Changsha 410082, China}

\author {Mingxing Chen}
\email{mxchen@hunnu.edu.cn}
\affiliation{Key Laboratory for Matter Microstructure and Function of Hunan Province,
Key Laboratory of Low-Dimensional Quantum Structures and Quantum Control of Ministry of Education, School of Physics and Electronics, Hunan Normal University, Changsha 410081, China}
\affiliation{State Key Laboratory of Powder Metallurgy, Central South University,  Changsha 410083, China}

\date{\today}

\begin{abstract}
In flat-band materials, the strong Coulomb interaction between electrons can lead to exotic physical phenomena. Recently, $\alpha$-In$_2$Se$_3$ thin films were found to possess ferroelectricity and flat bands. In this work, using first-principles calculations, we find that for the monolayer, there is a Weyl point at $\Gamma$ in the flat band, where the inclusion of the spin-orbit coupling opens a gap. Shifting the Fermi level into the spin-orbit gap gives rise to nontrivial band topology, which is preserved for the bilayer regardless of the interlayer polarization couplings. We further calculate the Chern number and edge states for both the monolayer and bilayer, for which the results suggest that they become quantum anomalous Hall insulators under appropriate dopings. Moreover, we find that the doping-induced magnetism for In$_2$Se$_3$ bilayer is strongly dependent on the interlayer polarization coupling. Therefore,  doping the flat bands in In$_2$Se$_3$ bilayer can also yield multiferroicity, where the magnetism is electrically tunable as the system transform between different polarization states.  Our study thus reveals that multiferroicity and nontrivial band topology can be unified into one material for designing multifunctional electronic devices.
\end{abstract}


\maketitle
Energy dispersions of electrons in crystals with a small bandwidth are called flat bands. 
Line-graph models such as Kagome and Libe lattices are found to host the flat bands, which can give rise to many exotic electronic properties, e.g., ferromagnetism, unconventional superconductivity, and nontrivial band topology.~\cite{Mielke1992,WuC2007,ShenR2010,Liu2014,leykam2018}  Over the past years, significant advances in this field have been made toward realizing the physics proposed for the lattice models in materials.  Experiments found that three-dimensional real materials such as Fe$_3$Sn$_2$~\cite{Ye2018,LinZ2018} and CoSn-type compounds~\cite{KangM2020NC,LiuZhong2020,Meier2020PRB,Kang2020NM,HanM2021NC} also possess flat bands. The reason is that the transition-metal atoms in the materials form a quasi-two-dimensional (2D) Kagome lattice.  Recently, thousands of flat-band materials were identified by high-throughput screening. ~\cite{Regnault2022,LiuH2021} The flat bands can also be obtained by structural engineering. One prominent way is to build moiré patterns using 2D materials such as graphene and transition-metal dichalcogenide monolayers by twisting one layer with respect to the other(s).~\cite{Bistritzer2011,CaoY1,CaoY2,Naik2018,Angeli2021,Xian1,Xian2} In such moiré systems, flat electronic bands are obtained by long-wavelength periodic potentials due to the large moiré unit cells, which folds and flattens the initial electronic bands of the material. It was demonstrated that doping the flat bands in these systems by a gate voltage, a phase diagram reminiscent of the cuprate superconductors could be obtained.~\cite{CaoY1} Interestingly, it was found that hole doping the flat bands in the MoTe$_2$-based moiré systems can give rise to  fractional quantum anomalous Hall effect.~\cite{ZengY2023,Anderson2023} 

Recently, In$_2$Se$_3$ monolayer was predicted to be a 2D ferroelectric material with out-of-plane (OP) polarization,~\cite{Ding2017} which was later confirmed by experiments.~\cite{Zhou2017,Cui2018,ZengH2018,ZengH2019}  In addition to its applications in ferroelectric devices, it has been widely used to manipulate electronic properties of surface layers in the heterostructures by making use of the switchable OP polarization.~\cite{Gong2019,Sun2020,Chen2020,Bai2020,Xue2020,Zhang2021,ZhaiB2021,
ChanY2021,FengD2021,ChengJ2022,SunR2022,DouK2022,ChenJ2023,Yu2023,ChaoJ2023,PanL2023} Moreover, electronic structure calculations find that the valence band (VB) is rather flat with a bandwidth in the order of hundreds of milli-electronvolt. For bilayers, density-functional theory calculations (DFT) reveal that twisting them by a large angle could lead to extremely flat bands (the bandwidth is only about 0.36 meV for a twisting angle of $9.43^\circ$) for the polarization state with tail-to-tail configuration.~\cite{LiC2021}  DFT calculations of In$_2$Se$_3$ monolayer find that hole doping leads to ferromagnetism with high Curie temperatures.~\cite{LiuChang2021,Duan2021} 

In this work, we investigate electronic and topological properties of In$_2$Se$_3$ monolayer and bilayer using first-principles calculations. We find that the In$_2$Se$_3$ monolayer is a normal insulator when the Fermi level is located in the band gap. Unexpectedly, we find that there is Weyl point in the flat valence band at $\Gamma$, where the spin-orbit coupling (SOC) opens a gap. Therefore, the system becomes topologically nontrivial as the Fermi level is shifted down to the spin-orbit (SO) gap.  Doping the flat band leads to ferromagnetism and chiral edge states, suggesting that quantum anomalous Hall effect can be obtained in this system.  Moreover, for the bilayer with hole doping,  the magnetism is strongly dependent on the interlayer polarization coupling, which allows for a ferroelectric tuning  of the magnetic properties. Our work suggest that doping the flat bands in In$_2$Se$_3$ thin films can induce multiferroicity and nontrivial band topology, of which the combination allows for developing multifunctional electronic devices.

The DFT calculations were performed using the Vienna Ab initio Simulation Package.~\cite{kresse1996} The pseudopotentials were constructed by the projector augmented wave method.~\cite{blochl1994,kresse1999} The exchange-correlation functional is parametrized using the Perdew–Burke–Ernzerhof (PBE) formalism under the generalized gradient approximation.~\cite{PBE1996} A 24 $\times$ 24 $\times$ 1 $\Gamma$-centered \textit{k}-mesh was used to sample the 2D Brillouin zone for structural relaxation and electronic structure calculations. An energy cutoff of 500 eV was used for the plane-waves for all calculations. A 20 \AA\ vacuum region was used between adjacent plates to avoid the artificial interaction between neighboring periodic images. Since the systems are layered systems, van der Waals (vdW) dispersion forces between the layers were accounted for through the DFT-D3.~\cite{DFT-D3} The structures were fully relaxed until the residual force on each atom was less than 0.01 eV/\AA. The topological properties calculations were carried out using the WANNIER90~\cite{wannier90} and WannierTools package.~\cite{WU2017}

\begin{figure}[htbp]
  \centering
  \includegraphics[width=.95\linewidth]{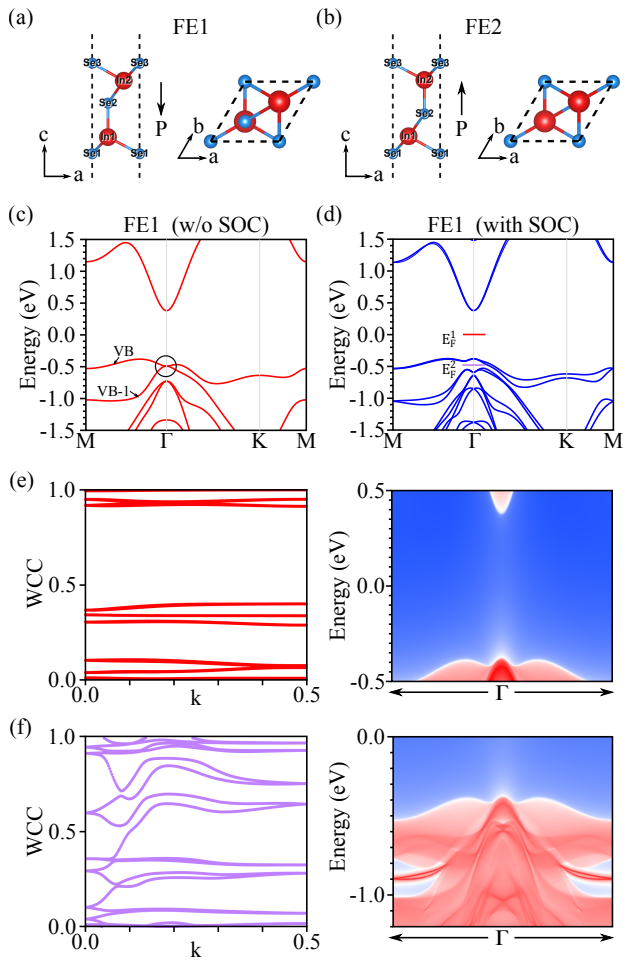}
  \caption{Geometric and electronic structures of In$_2$Se$_3$-1L. (a, b) The two polarization states of In$_2$Se$_3$-1L, which are referred to as FE1 and FE2, respectively. The character $P$ denotes the polarization and the arrows represent its directions. The atoms are labeled by numbers. (c) and (d) show the band structures without and with spin-orbit coupling, respectively. (e) and (f) show the WCC and edge states for In$_2$Se$_3$-1L as the Fermi level is placed at E$_F^1$ and E$_F^2$, respectively. }
  \label{fig1}
  \end{figure}

Figures~\ref{fig1}(a) and (b) show the geometric structure of the ferroelectric $\alpha$-In$_2$Se$_3$ monolayer (In$_2$Se$_3$-1L). There are three Se atoms and two In atoms in one unit cell.  The three Se atoms are named as Se1, Se2, and S3, respectively since they are inequivalent.  The In atoms are labeled as In1 and In2, respectively. The upward/downward motion of Se2 gives rise to down/up polarization.  The two polarization states are referred to as FE1 and FE2, respectively.  Figs.~\ref{fig1}(c) and d show its band structure without and with spin-orbit coupling (SOC).  The one without SOC shows that it is a semiconductor with a gap of about 0.8 eV, which is in agreement with previous studies.  The valence band is rather flat, whose bandwidth is less than 0.2 eV along $\Gamma$-M. Such flat bands allow for ferromagnetism by carrier doping.~\cite{LiuChang2021,Duan2021}  In addition, we note that the VB and the one lower than it cross at $\Gamma$ [marked by the circle in Fig.~\ref{fig1}(c)]. We find that the crossing point is a Weyl point.  These two states at $\Gamma$ belong to the $E$ representation of the $C_{3v}$ point group symmetry.  Using them as the basis,  an effective Hamiltonian up to the $k$-quadratic oder has the following form.

\begin{equation}
\begin{aligned}
H = &\scalebox{0.81}{$\begin{bmatrix}
A_1k_x+A_2(k_x^2+k_y^2)+A_3(k_x^2-k_y^2) & -A_1k_y+2A_3k_xk_y \\[0.3em]
-A_1k_y+2A_3k_xk_y & A_1k_x+A_2(k_x^2+k_y^2)-A_3(k_x^2-k_y^2)
\end{bmatrix}$}
\end{aligned}
\end{equation}


Upon including SOC, the profile of the conduction band remains unchanged. However, there is a gap opening at the crossing point, whose size is about 0.2 eV [see Fig.~\ref{fig1}(d)].  The profile of the bands near the SO gap at $\Gamma$ is similar to that for Bi$_2$Se$_3$,~\cite{ZhangH2009} which is a topological insulator.  We performed calculations of the evolution of the Wannier charge center (WCC) for In$_2$Se$_3$-1L using well fitted maximally-localised Wannier functions.  The results are summarized in Figs.~\ref{fig1}(e) and (f). We obtain Z2 = 0 when the Fermi level is located in the band gap between the valence and conduction bands [E$_F^1$ in Fig.~\ref{fig1}(d)].  By shifting the Fermi level into the SO gap [E$_F^2$ in Fig.~\ref{fig1}(d)],  we obtain Z2 = 1 [Fig.~\ref{fig1}(f)].  We further performed calculations of edge states for the system. Our results find that there are no gapless edge states for the Fermi level located in the band gap(E$_F^1$). Instead, when there are gapless edge states when the Fermi level is in the SO gap (see Fig.~\ref{fig1}(f) for E$_F^2$). Therefore, one can obtain nontrivial topological properties by shifting the Fermi level into the SOC gap, which may be achieved by hole dopings.

\begin{figure}[htbp]
  \centering
  \includegraphics[width=.95\linewidth]{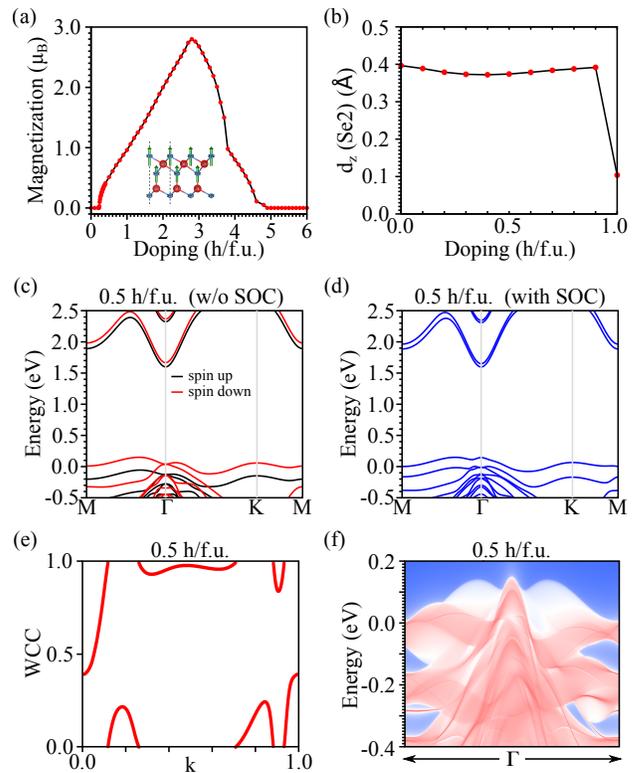}
  \caption{Doping effects on the geometric and electronic structures of In$_2$Se$_3$-1L. (a) The induced magnetization as a function of doping.  (b) The displacement of atom Se2 as a function of doping. (c) and (d) respectively show the band structures of In$_2$Se$_3$-1L at 0.5 h/f.c. without SOC and with SOC. (e) and (f) show the calculated WCC and edge states, respectively.}
  \label{fig2}
  \end{figure}

We now discuss the effects of hole doping on the electronic properties of In$_2$Se$_3$-1L. Note that the hole doping is to place the Fermi level at the flat bands, which may induce ferromagnetism as revealed by previous calculations.~\cite{LiuChang2021,Duan2021} Our calculations find that the system remains nonmagnetic upon dopings less than 0.23 holes per formula uint (h/f.u.). When the doping level exceeds this value, ferromagnetism is obtained and the magnetic moments increase sharply.  The magnetization is enhanced when the doping keeps increasing up to 3.0 h/f.u.. The inset of Fig.~\ref{fig2}(a) shows the magnetic moments are mainly contributed by the surface Se atoms.  Se2 and Se3 atoms have larger magnetizations than Se1 atom.  This effect can be understood since the valence band is dominated by contributions of Se2 and Se3.  Our structural relaxations find that the ferroelectricity is preserved for dopings until 1.0 h/f.u. . Our results are consistent with previous studies.~\cite{LiuChang2021,Duan2021}  

Figure~\ref{fig2}(c) shows the band structures without SOC for a doping of 0.5 h/f.u., from which one can clearly see spin splittings in the valence band. The splittings in the conduction band are less significant than those in the valence band. This effect can be understood since the Coulomb interaction between electrons in the flat bands play a dominating role in the magnetism.  Like the undoped case, there remains a band crossing at the $\Gamma$ point near the Fermi level.  Inclusion of SOC leads to band gap opening at the crossing point [see Fig.~\ref{fig2}(d)]. We investigate its topological properties by calculating the Chern number and edge states, of which the results are shown in Figs.~\ref{fig2}(e) and (f).  We obtained $C$ = 1 for the doping of 0.5 h/f.u..  We have also performed calculation for other hole dopings, at which ferromagnetism appears. Our results show that the system has $C$ = 1 along as the Fermi level is located in the SO gap. The number of chiral edge states changes from two for the undoped system to one for the doped system. These results suggest that  the hole-doped In$_2$Se$_3$-1L may be the platform for observing the quantum anomalous Hall effect.   
%
There are four polarization configurations for an In$_2$Se$_3$ bilayer (In$_2$Se$_3$-2L), which are dependent on the interlayer polarization couplings.~\cite{Ding2017} Two have FE interlayer couplings with opposite polarizations and the other two have antiferroelectric (AFE) couplings.  However, there is still a difference between the AFE states,  that is, the polarizations in them are in tail-to-tail and head-to-head orderings, respectively.  Our calculations find that the state with tail-to-tail interlayer couplings has the lowest energy among all the states.  The trend in the relative stability remains unchanged upon hole dopings less than 0.6 h/f.u.. The AFE states are insulators with the band gap of about 0.6 and 0.4 eV for C2 and C3, respectively. Whereas the FE states are semimetals.  The reason is that the built-in electric field shifts the bands of the two layers. As a result, for the state C1 with down polarizations, the VB of the top layer and the CB of the bottom layer overlap. On the other hand, the atoms contributed to these two bands are far away from each other (the VB is contributed by Se6 and the CB is mainly contributed by Se1). Therefore, the band gap opening induced by the hybridizations between their orbitals are small.  In fact, the two bands intersect at a $k$-point along $\Gamma$-K and a tiny band gap occurs along $\Gamma$-M.  

  \begin{figure}[htbp]
  \centering
  \includegraphics[width=.95\linewidth]{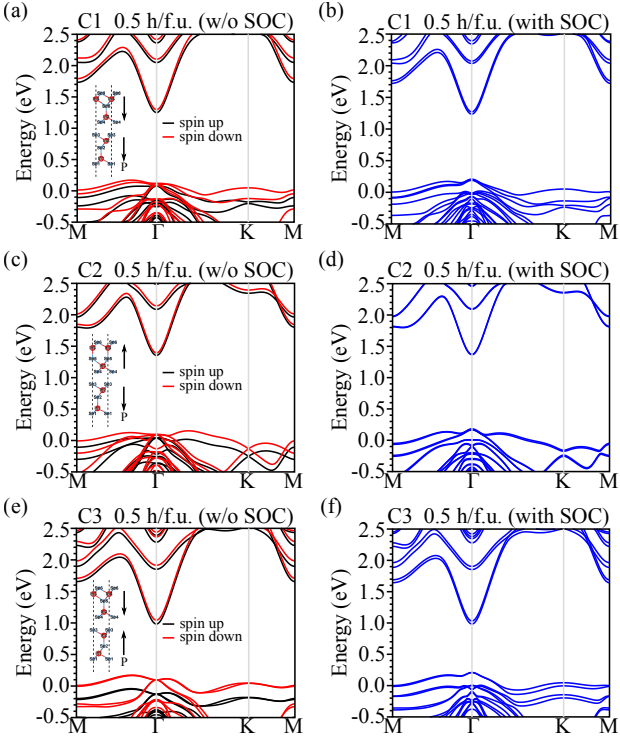}
  \caption{Band structures for three polarization states without and with SOC. The polarizations in the layers are denoted by arrows. The insets depict the the polarization states, which are named C1, C2, and C3, respectively. (a, b), (c, d), and (e, f) are for C1, C2, and C3, respectively.}
  \label{fig3}
  \end{figure}

Similar to In$_2$Se$_3$-1L, a low concentration of hole doping does not yield ferromagnetism. Increasing the doping level can give rise to ferromagnetism for all the states since the flat bands near the Fermi level still exist. Furthermore, the state with both intra- and inter-layer FM couplings is more stable than others. Figs.~\ref{fig3}(a) and (b) shows the band structure for the FE state C1 under a doping level of 0.50 h/f.u. . One can see that in addition to the spin splitting, doping also enhances the gap between the valence and conduction bands. This effect is due to that the doped carriers screen the built-in electric field. We investigate the doping effect on the difference in potential between the two surfaces of C1. Our results find that the potential difference decreases monotonically as a function of doping (not shown), which confirms that the doping reduces the built-in electric field.   Correspondingly, the relative shifting of the bands for the two layers is shrinked, which enhances the band gap correspondingly. For the two AFE states, similar spin splittings happen to the VB. A comparison of the band structures for all the states finds that the band splitting in the VB is the largest for C2 and is the smallest for C3, which gives rise to the same trend in the net magnetization as will be discussed below. 

\begin{figure}[htbp]
  \centering
  \includegraphics[width=.95\linewidth]{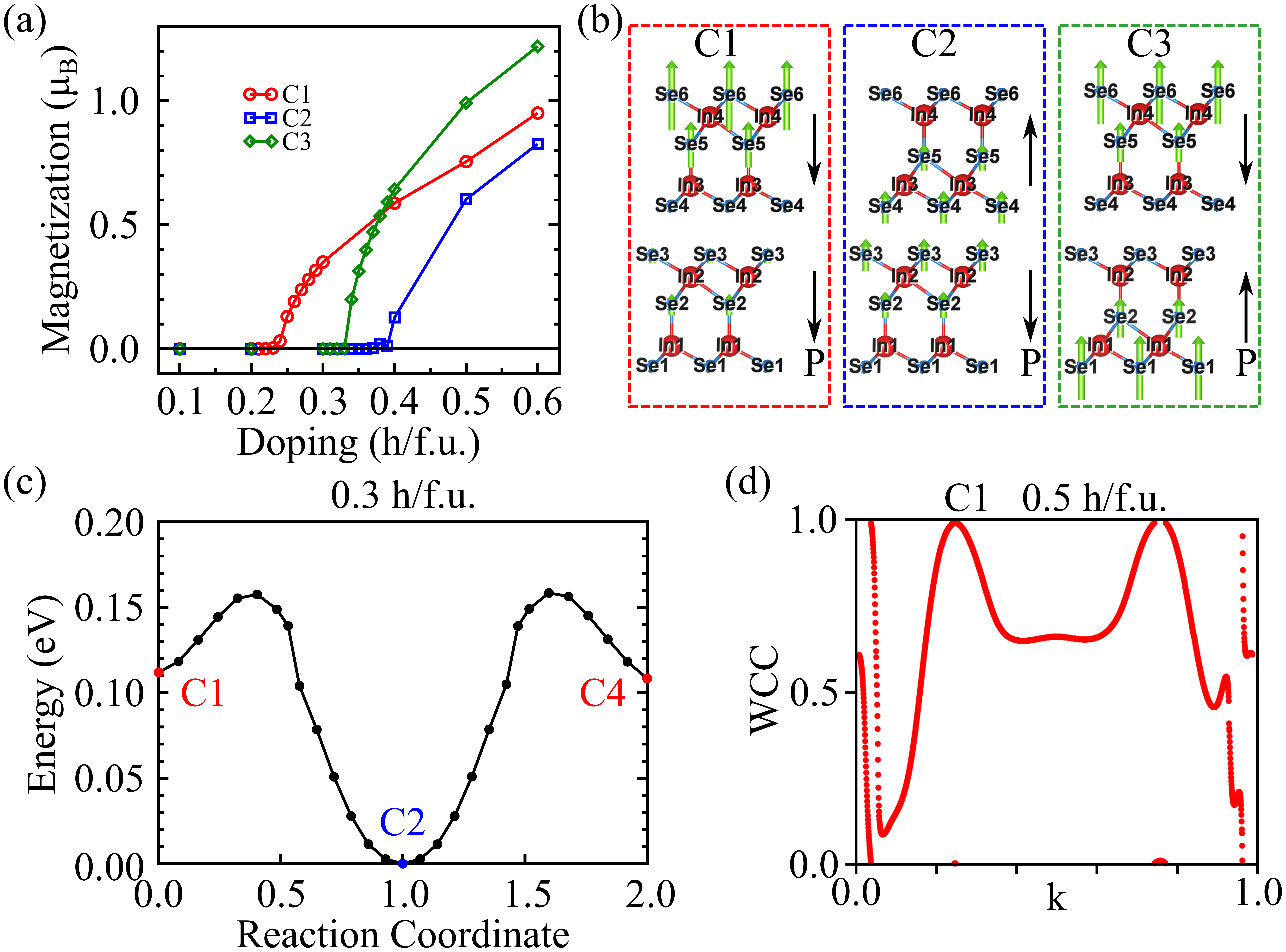}
  \caption{Polarization-depdendent ferromagnetism in hole-doped In$_2$Se$_3$-2L.  (a) Magnetization as a function of hole doping. (b) Distribution of magnetization on atoms for C1, C2, and C3, respectively. The arrows denote the polarization in the layers. (c) Kinetic pathways for the transforming of polarization states of In$_2$Se$_3$-2L with a doping of 0.3 h/f.u.. (d) WCC for C1.}
  \label{fig4}
  \end{figure}

The magnetization of In$_2$Se$_3$-2L shows a strong polarization-dependence. Take the doping level of 0.3 h/f.u. as an example, only configuration C1 becomes ferromagnetic at this doping level, for which the net magnetization is about 0.4 $\mu_B$ [Fig.~\ref{fig4}(a)].  Whereas C2 and C3 remain to be nonmagnetic. Ferromagnetism emerges for C2 and C3 by increasing the doping level. However, there are differences in the net magnetization between them. For instance, at 0.50 h/f.u. , the magnetizations are about 0.75, 0.60, and 0.99 $\mu_B$ for C1, C2, and C3, respectively. Moreover, distribution of the magnetic moments are also polarization-dependent. Figure ~\ref{fig4}(b) shows the magnetic moments on the atoms for the considered states. For C1, i.e., the FE state, the magnetizations are confined to top layer, in which the surface atom Se6 has the largest value.  For the other FE state that has opposite polarization to C1, the magnetic moment on Se6 is vanishing and that on Se1 dominates the contributions to the magnetization.  For C3, the Se atoms on the two surfaces have much larger magnetic moments than other atoms. Whereas for C2, the magnetization is mainly distributed on the interlayer Se atoms.

The polarization-dependence of the induced ferromagnetism allows for ferroelectric tuning of magnetism for In$_2$Se$_3$-2L.  Because the AFE state C2 is the ground state, the In$_2$Se$_3$-2L has a layer-by-layer flipping [Fig.~\ref{fig4}(c)]. Suppose that the system is in the FE state with configuration C1 (the polarization pointing downward), which can be achieved by external electric field anyway.  An appropriate upward electric field that is sufficiently large to flip the polarization of the top layer then drives the system transforming from C1 to C2. As a result, the system changes from nonmagnetic to ferromagnetic for the doping in the range of 0.25 h/f.u. and 0.33 h/f.u.. At 0.5 h/f.u., the net magnetization changes from 0.75  $\mu_B$  to 0.60 $\mu_B$. Further enhancing the electric field can eventually have the system transform into another FE state with upward polarization. This operation again leads to a change in the magnetism. Therefore, the doped In$_2$Se$_3$-2L can be a multiferroic system with electrically tunable magnetism.

Moreover, we performed SOC calculations for C1, C2, and C3 with various hole dopings. Our results show that all have the same behavior as the doped In$_2$Se$_3$-1L, that is, a SO gap is obtained in the VB at $\Gamma$. We further performed calculations of Chern number for all the considered states, which reveal that they all have nonzero Chern number when the Fermi level is placed in the SO gap [Fig.~\ref{fig4}(d)]. Calculations of edge states for a semi-infinite ribbon also find that there is a gapless band in the SOC gap (not shown), which suggests that In$_2$Se$_3$-2L can be a quantum anomalous Hall insulator by appropriate hole dopings.

  In conclusion, we have investigated electronic and topological properties of In$_2$Se$_3$-1L and In$_2$Se$_3$-2L under hole doping using first-principles calculations. We find that the In$_2$Se$_3$-1L is a normal insulator with flat valence band. However, there is a Weyl point in the flat band, where inclusion of SOC opens a gap of about 0.2 eV. We find that nontrivial topological properties can be obtained when the Fermi level is shifted into the SO gap, which can be achieved by hole doping. Moreover, our results reveal that doping the flat band could induce ferromagnetism with C = 1 and chiral gapless edge states. For In$_2$Se$_3$-2L, we find that doping the flat bands not only lead to nontrivial topological properties, but also induces polarization-dependent ferromagnetism.  Our results suggest that In$_2$Se$_3$ thin films provide the platform for realizing quantum anomalous Hall effect and multiferroicity, which can be further used to design multifunctional electronic devices.

\begin{acknowledgments}
This work was supported by the National Natural Science Foundation of China (Grants No. 12174098, No. 11774084, No. U19A2090, No. 12204378, and No. 52372260) and Project supported by State Key Laboratory of Powder Metallurgy, Central South University, Changsha, China. Calculations were carried out in part using computing resources at the High Performance Computing Platform of Hunan Normal University.
\end{acknowledgments}

\bibliography{refs}
\bibliographystyle{apsrev4-2}

\end{document}